\documentclass[twocolumn,twoside,slac_two]{revtex4}
\usepackage{graphicx}
\usepackage{fancyhdr}
\begin{document}


\title{Stochastic Particle Acceleration in Parallel Relativistic Shocks}


\author{Joni J. P. Virtanen}
\affiliation{Tuorla Observatory, V\"ais\"al\"antie 20, 
  FI-21500 Piikki\"o, FINLAND}
\author{Rami Vainio}
\affiliation{Department of Physical Sciences, 
  P.O. Box 64, FI-00014 University of Helsinki, FINLAND}


\begin{abstract}
 We present results of test-particle simulations on both the first-
 and the second-order Fermi acceleration for relativistic parallel
 shock waves. Our studies suggest that the role of the second-order
 mechanism in the turbulent downstream of a relativistic shock may
 have been underestimated in the past, and that the stochastic
 mechanism may have significant effects on the form of the particle
 spectra and its time evolution.
 Figures are reproduced from Virtanen \& Vainio 2005 \cite{VV2005} 
 by the permission of the AAS.
\end{abstract}


\maketitle
\thispagestyle{fancy}


\section{INTRODUCTION}

Particle acceleration in relativistic shocks is typically considered 
to be due to either the first- or the second-order Fermi acceleration 
(stochastic acceleration). 
The former of these acts at the shock front where charged particles 
gain energy by subsequently scattering on different sides of the shock
front, whereas for the latter case the velocity difference between the
scattering centers is provided by the turbulence alone.
The first-order mechanism is well known to produce power-law particle 
energy spectrum $N(E)\propto E^{-\sigma}$ with spectral index $\sigma$ 
depending on compression ratio $r$ of the shock as $\sigma = (r+2)/(r-1)$ 
for nonrelativistic speeds \citep[e.g.,][]{Drury1983}, 
and approaching to value $\sigma\approx2.2$ as the speed increases 
to ultrarelativistic, almost regardless of the other shock conditions 
\citep[e.g.,][]{KirkEtAl2000,AchterbergEtAl2001,LP2003,VV2003a}.

However, problems arise with flatter spectra ($\sigma\lesssim2$); 
although the first-order mechanism can produce harder spectra with 
$\sigma\to1$ depending, for instance, on the injection model \cite{VV2003b}
or the scattering center compression ratio 
\citep[e.g.][]{EllisonEtAl1990,VVS,VVS2005}, it is not able to produce 
spectral indices flatter than $\sigma=1$. 

Stochastic acceleration, on the other hand, has been known to be 
present in the turbulent downstream of shocks, but mostly because 
it works on much longer timescales than the almost instantaneous 
first-order mechanism \citep[e.g.,][]{CS1992,VS1998} it has been been
neglected for most of the cases (note, however, e.g. \cite{Ostrowski1993}).
However, for non-thermal particle distributions radiating in
astrophysical objects the bulk of radiation is emitted by the 
particles that have already left the
shock front towards the downstream. Thus, the second order mechanism
has, indeed, more time available to for acceleration than the first-order
process. So while the neglecting of the stochastic process in calculation
of the accelerated particle spectra right at the shock front could be
justified, it is not possible to neglect its effect on the spectrum in
shocks in general, especially when basically all of the astrophysical 
sources related to relativistic shocks, remain still spatially unresolved.

Here we study the possibility of the stochastic process to
have visible effects on the particle energy spectrum in parallel 
relativistic shock waves in two different cases: (i)
particles injected at the shock front and accelerated further by
the first order mechanism and (ii) particles drawn from the heated
(but not shock accelerated) particle population of the downstream
region of the shock.
We apply numerical Monte Carlo test-particle
simulations and focus on shocks that, in addition to being parallel, have
small-to-intermediate Alfv\'enic Mach number. Low Mach numbers for 
relativistic shocks could prevail in magnetically dominated jets that
are lighter than their surroundings, e.g. in pair-plasma jets.

\section{MODEL}

The applied model, together with its numerical implementation, is 
described in detail in ref.~\cite{VV2005}, 
and here only the essential parts of the model are explained.

We use the shock rest frame as our basic coordinate system. In the case
of parallel shock the flow direction -- as well as the large scale magnetic 
field -- is perpendicular to the shock normal in this frame.
In this frame the shock lies at the origin, negative values of location $x$ 
mean the upstream and positive values the downstream.
The plasma frame, i.e. the frame where the bulk of the plasma is at
rest, is moving with the local flow speed $V$ with respect to the 
shock frame. 
Waves propagate at phase speed $V_\phi$ in the plasma frame in directions
both parallel and antiparallel to the flow, so in the shock frame waves 
move at speed $(V+V_\phi)/(1+VV_\phi/c^2)$ ($c$ is the speed of 
light), and the frame moving at this speed (i.e. flowing with 
the waves) is called the wave frame or the rest frame of the 
scattering centers.
If the scattering centers are taken to be fluctuations 
frozen-in to the plasma then the speed of the
waves with respect to the underlying flow is $V_\phi=0$ and the
plasma frame is also the rest frame of the scattering centers.
Here we assume the waves to be Alfv\'en waves, and the phase speed 
to be the Alfv\'en speed
\begin{equation}
  v_{\rm A}=\frac{B_{0} c}{\sqrt{4\pi hn + B_{0}^{2}}},
  \label{eq:alfven_speed}
\end{equation}
where $B_0$ is the large scale magnetic field, and $h$ and $n$ the 
specific enthalpy and the number density, all in the local plasma frame.

We set the proper speed of the shock to be $u_1 \equiv \Gamma_1 V_1 = 10 c$
(corresponding to shock speed $V_1 \approx 0.995 \, c$), and model the shock 
transition with a very thin hyperbolic tangent profile 
\cite{SK1989} with thickness $W \approx \frac{1}{100}$th of the 
mean free path of an upstream particle. For this width the shock 
can still be considered almost step-like \citep{VV2003a}.
\begin{figure}
\includegraphics[width=0.95\linewidth]{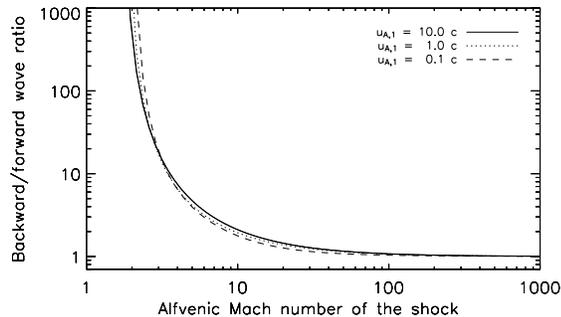}
\caption{Ratio of the amplified wave intensities 
as function of Alfv\'enic Mach number 
for three proper Alfv\'en speeds. Here $q=5/3$. \cite{VV2005} }
\label{fig:amplifications}
\end{figure}

We model the downstream turbulence as a superposition of Alf\'en waves
propagating parallel and anti-parallel to the plasma flow at constant
speed. The turbulence spectrum has a power-law form with spectral
index $q$ for wavenumbers above some inverse correlation length $k_0$.
For $k < k_{0}$ the wave intensity per logarithmic bandwidth is
assumed to be equal to the background field intensity, i.e.,
$I(k)=B_0^2 k^{-1}$ for $k<k_{0}$. In this work we use two values for
$q$: $2$, producing rigidity independent mean free paths, and $5/3$,
being consistent with the Kolmogorov phenomenology of turbulence.

The downstream wave intensities can be calculated from known upstream
parameters \citep[e.g.,][]{VS1998,VVS,VVS2005}, and regardless of the
cross helicity of the upstream wave field (only parallel or
anti-parallel waves, or both), there are always both wave modes
present in the downstream region; in this work we assume the cross helicity
to vanish in the upstream.  The amplification of the waves at
the shock depends on the strength of the magnetic field as well as on
the form of the turbulence spectrum, as shown in
Fig.~\ref{fig:amplifications} where the amplification of the different
wave modes is shown as a function of (quasi-Newtonian) Alfv\'enic Mach
number
\begin{equation}
  M = u_{\rm A,1} / u_1.
\end{equation}
\begin{equation}
  u_{{\rm A,1}}=v_{{\rm A,1}}/\sqrt{1-\beta_{{\rm A,1}}^{2}},
\end{equation}
is the upstream Alfv\'en proper speed with 
$\beta_{{\rm A,1}}=v_{{\rm A,1}}/c$.
The waves are seen to propagate predominantly backward for relatively 
low-Mach-number shocks; this is the case for both the relativistic 
\cite{VVS} and the nonrelativistic \cite{VS1998} shocks. This enables the
scattering center compression ratio $r_{k}$ to grow larger than the
gas compression ratio $r$ and, thus, to cause significantly harder
particle spectra compared to the predictions of theories relying on
fluctuations frozen-in to plasma flow \cite{VVS,VVS2005}. 
As the Mach number increases, the downstream wave intensities 
approach equipartition at the ultra-relativistic limit.


Electrons scatter off the magnetic fluctuations resonantly. The
scattering frequency of electrons with Lorentz factor $\gamma$ is
determined (see ref. \cite{VV2005} for details) by the intensity of 
waves at the resonant wavenumber 
\begin{equation}
  k_{{\rm res}} = \frac{\Omega_{{\rm e}}}{v}
  = \frac{\Omega_{{\rm e},0}}{c\sqrt{\gamma^2-1}},
  \label{eq:resonant_wavenumber}
\end{equation}
where
\(  \Omega_{{\rm e}}=\frac{\Omega_{{\rm e,0}}}{\gamma} \)
is the relativistic electron gyrofrequency and $\Omega_{\rm e,0} =
\frac{eB}{m_{{\rm e}}c}$ is its non-relativistic counterpart.

Scatterings are elastic in the wave frame and the existence of waves
propagating in both directions at a given position, thus, leads to
stochastic acceleration \citep{Schlickeiser1989}.  Since the spectrum
of waves is harder below $k=k_0$, scattering at energies
\begin{equation}
  \gamma > \gamma_{0} \equiv \frac{\Omega_{{\rm e,0}}}{k_{0}c}\gg1
  \label{eq:gamma_0}
\end{equation}
becomes less efficient. This suggests that the electron acceleration
efficiency should decrease at $\gamma > \gamma_0$. Instead of trying to
fix the value of $k_0$, we use a constant value $\gamma_{0}=10^{6}$,
which is in agreement with observations of maximum Lorentz factor of
electrons in some AGN jets \citep{MeisenheimerEtAl1996}.

In addition to scattering, the particles are also assumed to lose
energy via the synchrotron emission.

\section{RESULTS \& DISCUSSION}

%
%
\begin{figure*}
  \begin{center}
    \includegraphics[width=0.46\linewidth]{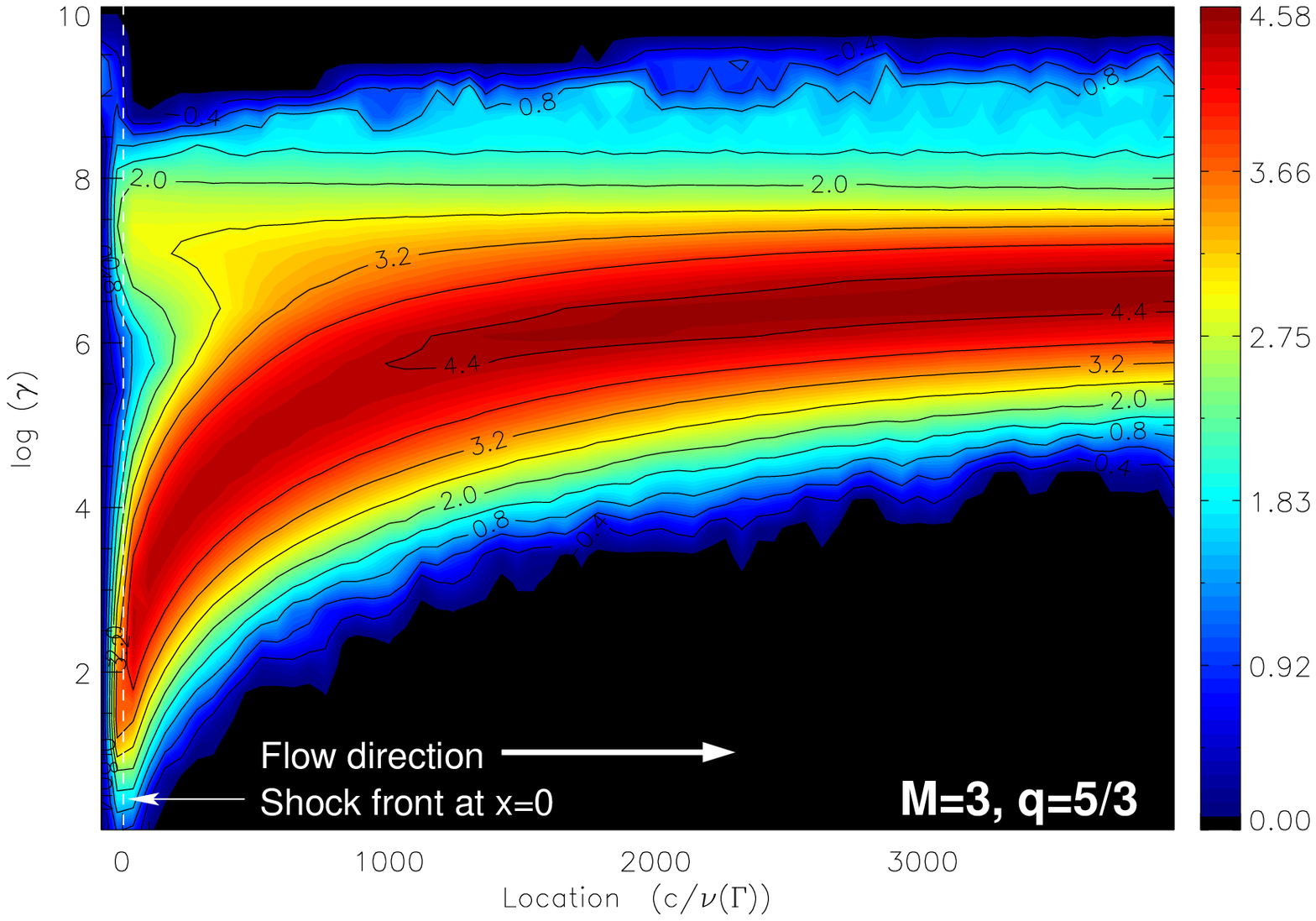}
    \includegraphics[width=0.46\linewidth]{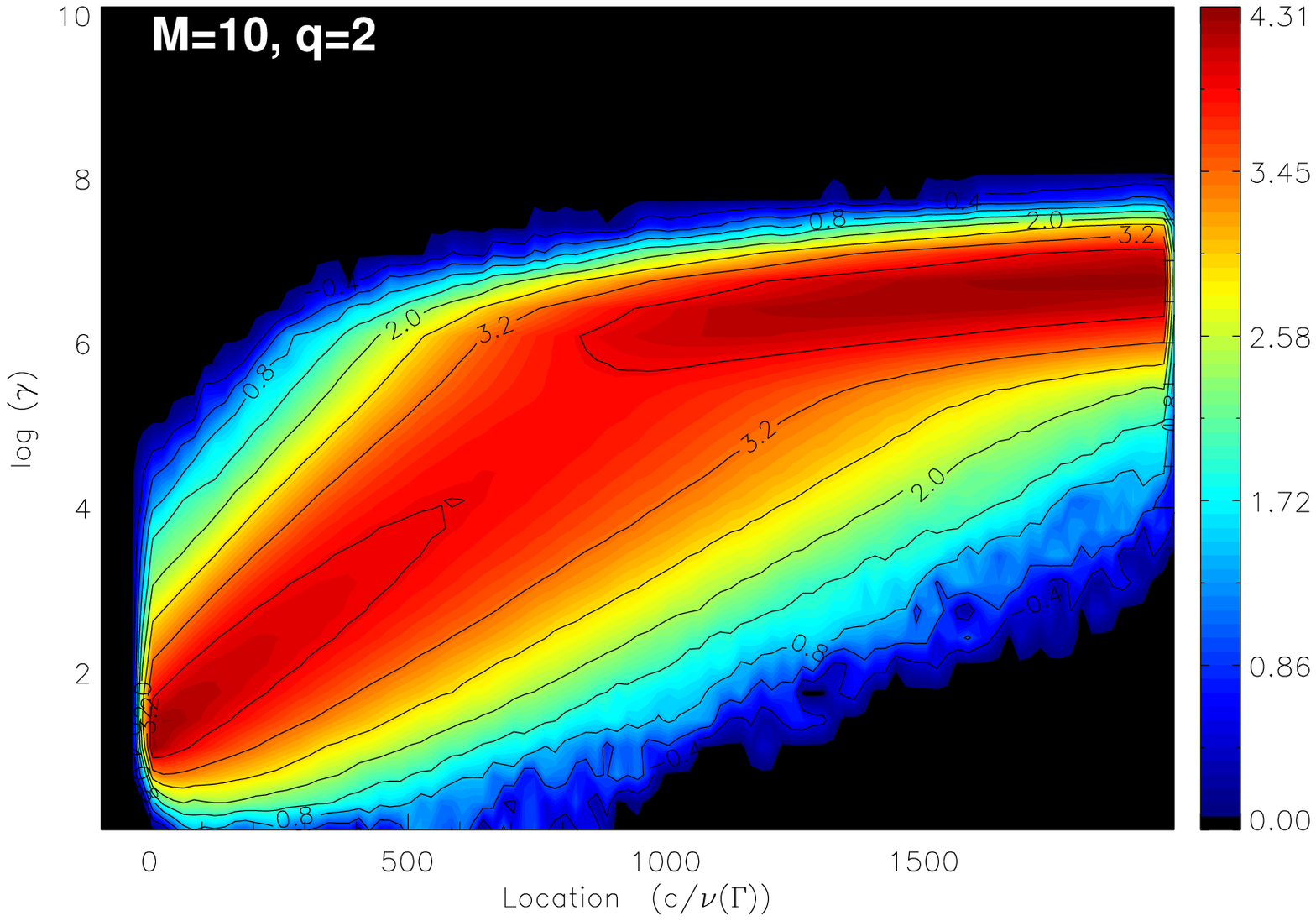}\\
    \includegraphics[width=0.45\linewidth]{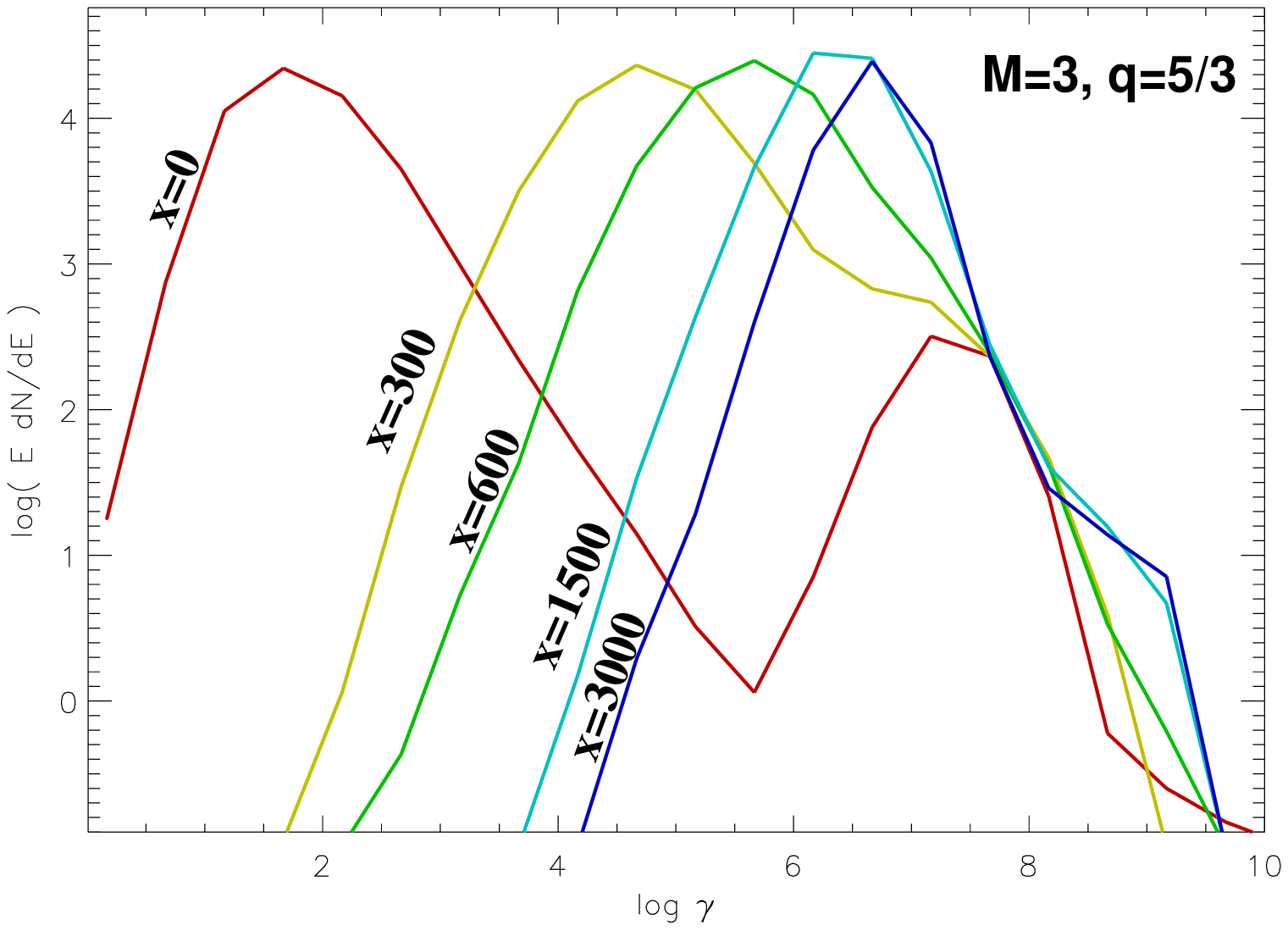}
    \includegraphics[width=0.45\linewidth]{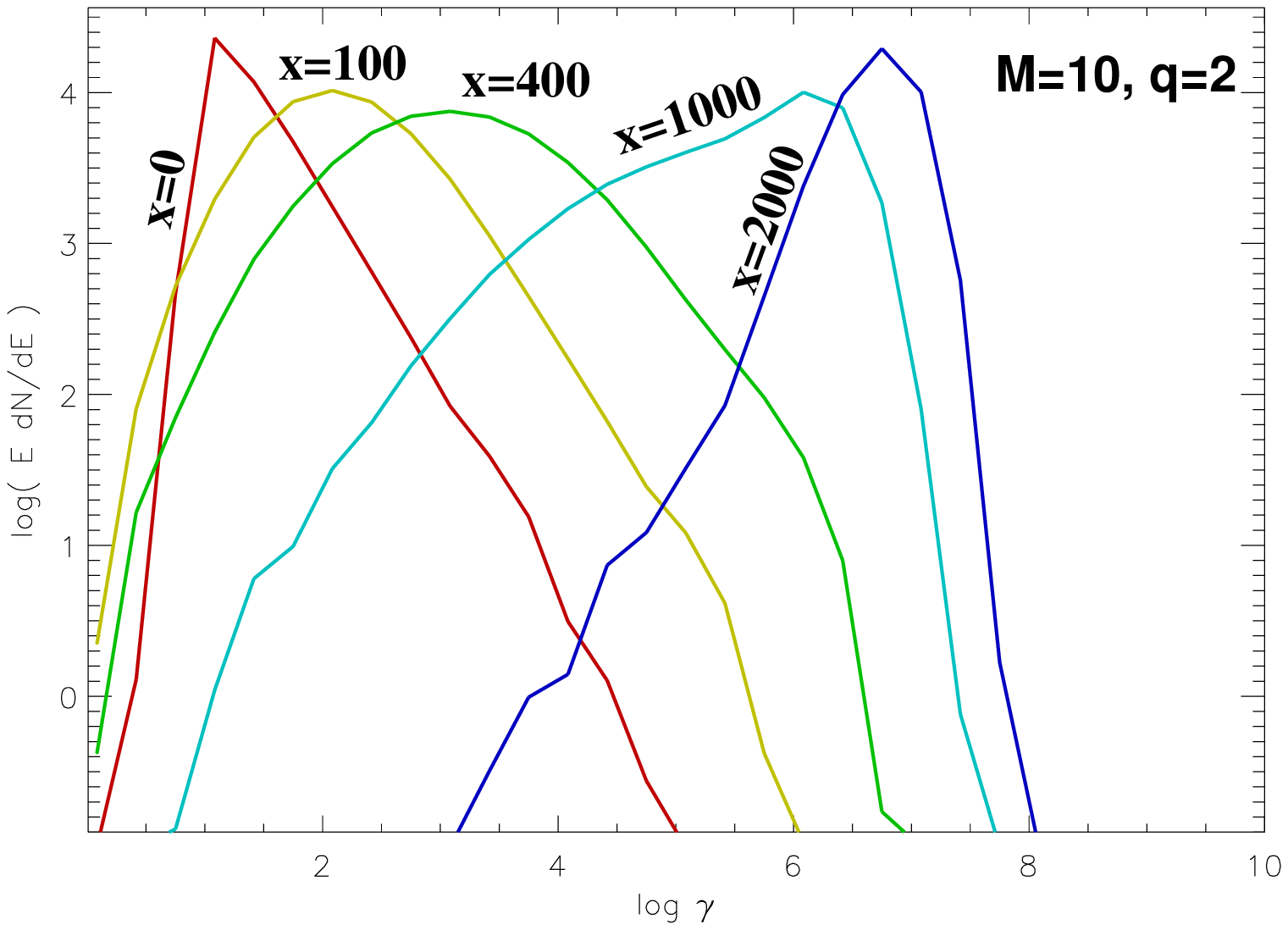}
  \end{center}
  \caption{Steady-state energy distribution of shock-injected particles
    at different distances from the shock. Upper panels show contours
    of \( \log(E\frac{dN}{dE}) \) as function of energy and location,
    while the lower panels present the energy distributions as slices
    at certain locations in the downstream.  On the left hand panel is
    a case with a shock with Alfv\'enic Mach number $M=3$ and for
    turbulence corresponding to the Kolmogorov turbulence, and on the
    right panel the same for $M=10$ and turbulence corresponding to
    that created by the particles themselves. Equipartition is assumed
    between forward and backward waves. \cite{VV2005}.}
\label{fig:contours_1}
\end{figure*}

Simulations were run separately for low, intermediate and high
Alfv\'enic-Mach-number shocks ($M=3$, $M=10,$ and $M=1000$,
respectively -- see Fig.~\ref{fig:amplifications} for the
corresponding wave intensity ratios), and for four cases of downstream
turbulence; for turbulence spectral index $q=2$ and $q=5/3$ with
downstream wave field calculated using wave transmission analysis
described earlier, and with the downstream forward and backward waves
being in equipartition. The proper speed of the shock is set to $u_1 =
10\, c$ in all simulations.

Some of the results (including all of the case of $M=1000$ 
for which the effects were, expectedly, barely visible, and 
all those for which the downstream turbulence was calculated using the
Alfv\'en wave transmission model) 
have been omitted from this paper due to limited space; 
all of them are, however, available at 
\url{http://www.astro.utu.fi/red/qshock.html}.


\subsection{Particles Injected at the Shock}

In the first case we studied the effect of stochastic acceleration on
particles that have been already accelerated at the shock. We injected 
the particles with relatively small initial energy into the simulation
right behind the shock and allowed them to continue accelerating via 
the stochastic process. This kind of injection simulates the case of some
already-energized downstream particles returning into the shock, but
removes need of processing the time consuming bulk of
non-accelerating thermal particles. The high-energy part of the
particle energy distribution -- which we are interested in in this
study -- is similar, regardless of the injection energy.

In the case of high Alfv\'enic Mach number ($M=1000$, corresponding to
magnetic field $B_0 \simeq 1.4$ mG in a hydrogen plasma and $\simeq 46
\, \mu$G in a pair plasma) the contribution of the stochastic process
to the energy distribution of the particles was, expectedly, very
insignificant compared to that of the first-order acceleration at the
shock. This was the case regardless of the applied turbulence properties, 
and because the wave intensities calculated from the Alfv\'en wave 
transmission analysis are very close to equipartition for 
high-Mach-number shock (see Fig. \ref{fig:amplifications}),
the difference to the case of explicitely assumed equipartition 
was, as expected, minimal.

For shocks with Alfv\'enic Mach number sufficiently
low ($M=10$ and $M=3$, corresponding to $\approx 0.1$--$0.5$~G in a
hydrogen plasma, and to $\approx 5$--$15$~mG in a pair plasma) the
effect was, on the contrary to the high-$M$ case, very pronounced.  
The stochastic process began to re-accelerate particles immediately 
after the shock front, and the whole spectrum started to slowly shift 
to higher energies in all kinds of turbulence models studied. 
Two examples for the
case assuming the downstream wave intensities to be in equipartition 
are shown in Fig.~\ref{fig:contours_1}.
%
%
\begin{figure*}
  \begin{center}
    \includegraphics[width=0.46\linewidth]{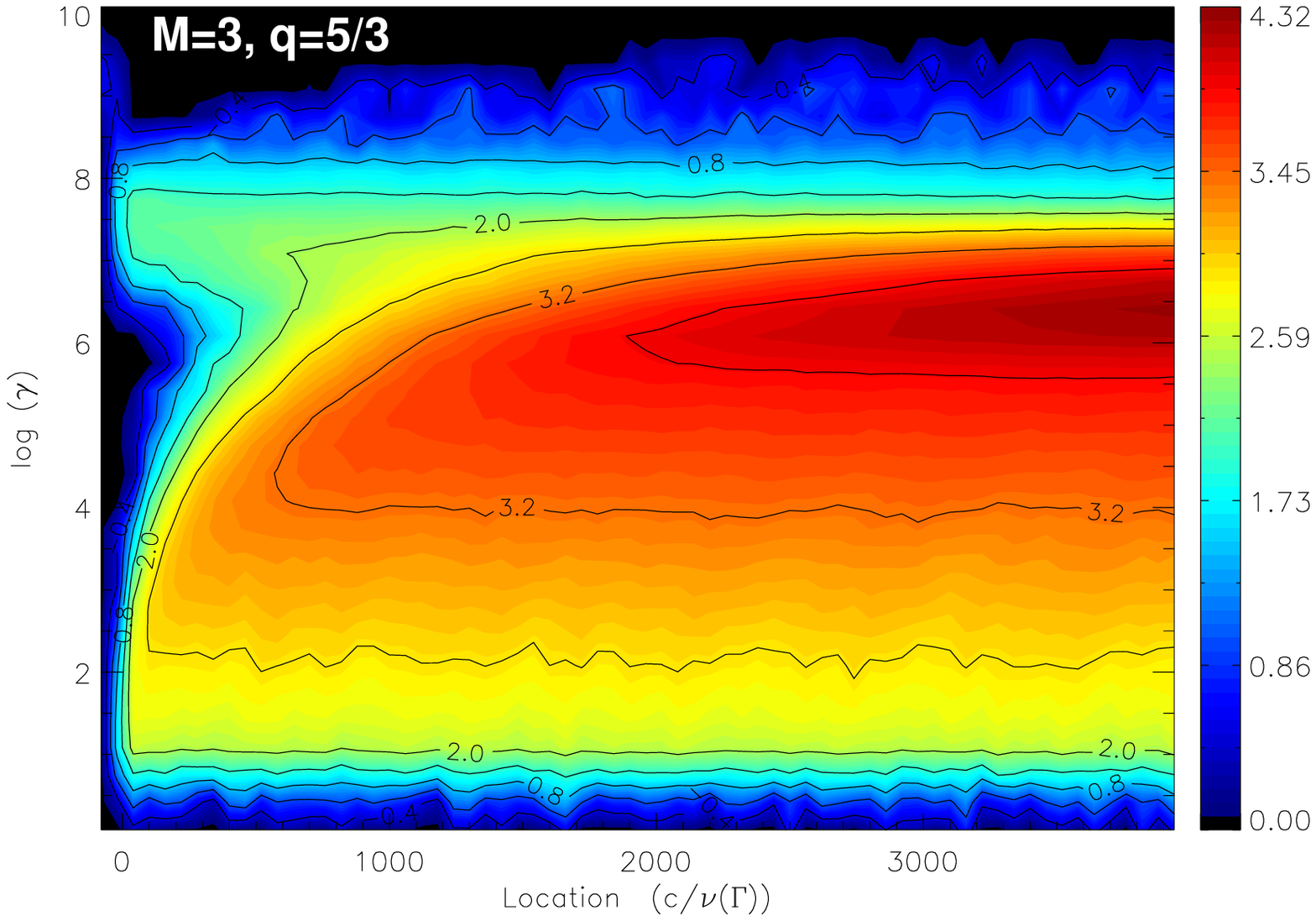}
    \includegraphics[width=0.46\linewidth]{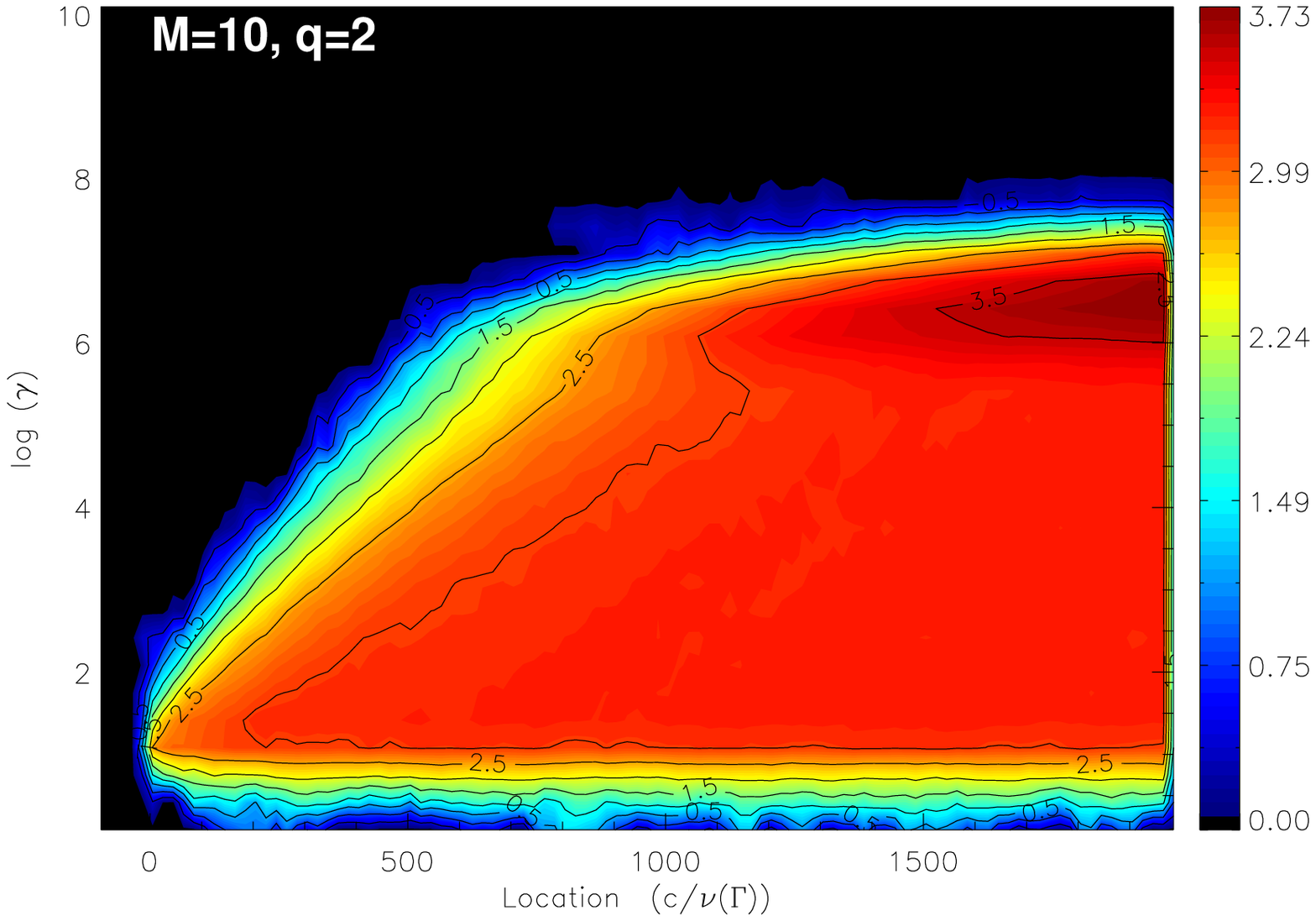} \\
    \includegraphics[width=0.45\linewidth]{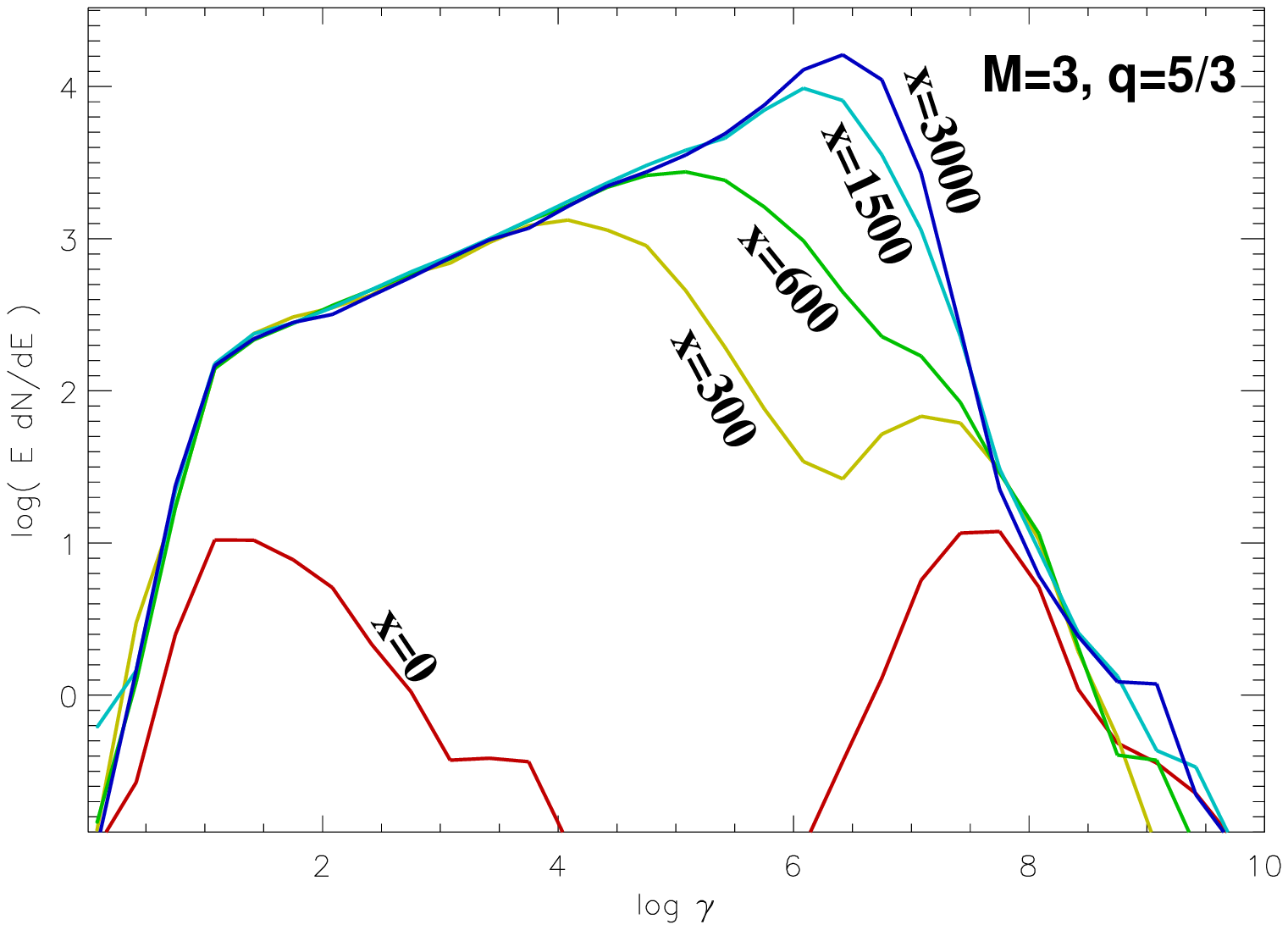}
    \includegraphics[width=0.45\linewidth]{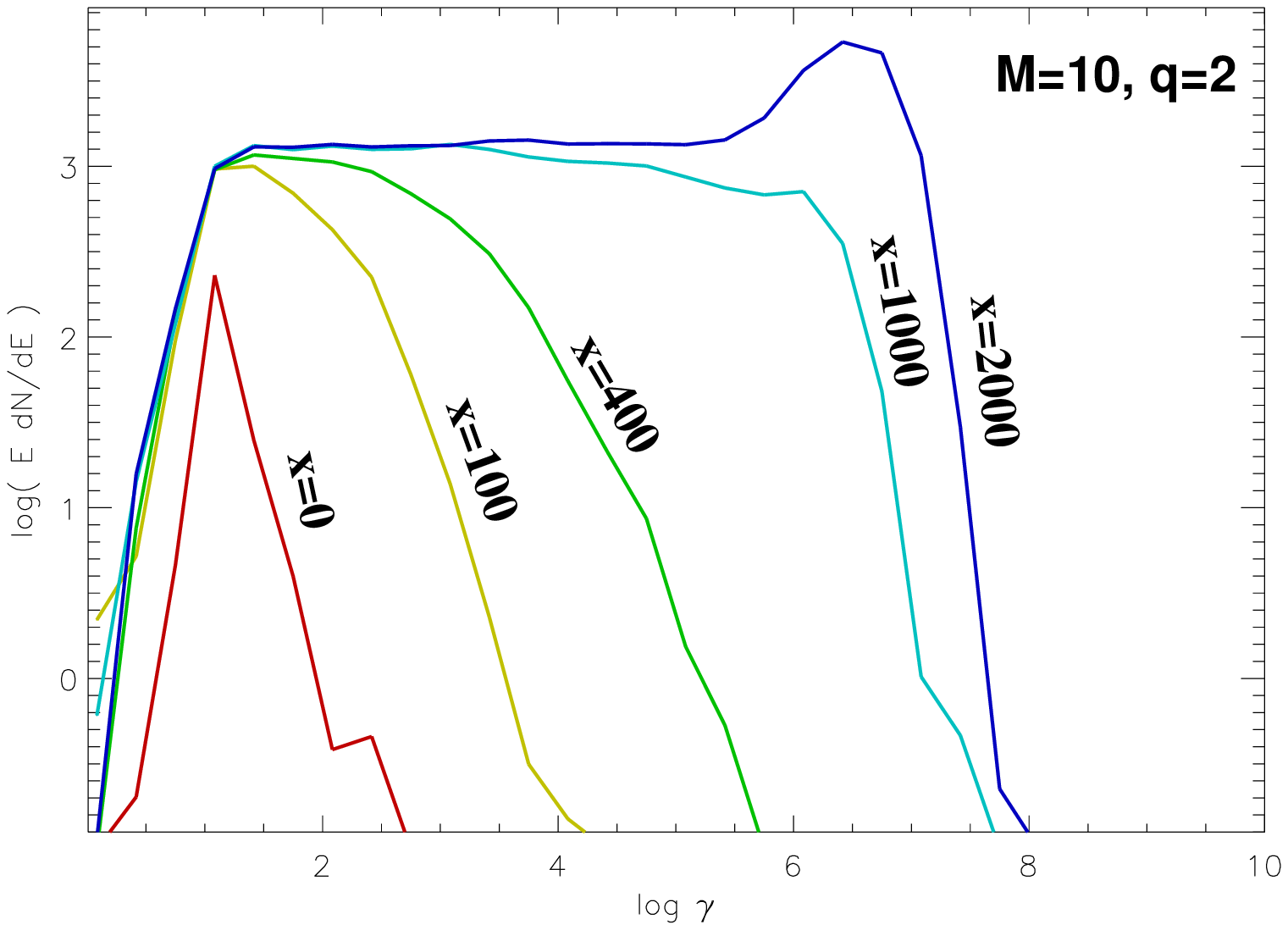}
  \end{center}
  \caption{Same as Fig.~\ref{fig:contours_1} but for particles 
    injected uniformly throughout the downstream region. 
    \cite{VV2005}.}
  \label{fig:contours_2}
\end{figure*}

When comparing otherwise similar cases that differ only for the
downstream cross helicity (i.e., whether the wave field is resulting
from the wave transmission calculations \cite{VVS} or an equipartition 
of parallel and anti-parallel waves is assumed), 
the calculated wave-transmission cases with more
anti-parallel waves (see \citep{VVS} and Fig.\ref{fig:amplifications})
show stronger first-order acceleration, but weaker stochastic
acceleration. This is because of the larger scattering center
compression ratio in the wave-transmission case leading to more
efficient first-order acceleration \citep[][and Virtanen \& Vainio, in
preparation]{VVS} and, on the other hand, faster momentum diffusion
rate in the equipartition case leading to more efficient stochastic
acceleration.


\subsection{Particles Injected Across the Downstream}

Next we assumed that a constant injection mechanism exists
throughout the downstream region -- this mimics a case where turbulent
fluctuations cascade to higher wavenumbers and inject a fraction
of the thermal electrons to the stochastic acceleration process. We
injected particles at a constant initial energy uniformly and
isotropically within the whole downstream region. 
Examples of results are shown in Fig.~\ref{fig:contours_2} for 
parameters otherwise equal to those in Fig.~\ref{fig:contours_1}.

The general behavior of the acceleration process is very similar for
both injection methods: the piling up of the particles with
$\gamma > \gamma_0$ is visible and the effect of the turbulence
spectrum is the same.  For the constant injection, however, particles
begin to form a power-law plateau in the energy range between
the injection energy and $\gamma_0$. The produced spectral index 
depends on the spectral index of the magnetic field fluctuations $q$ 
as $\sigma = q - 1$ \cite{VV2005, Schlickeiser1989}; 
for $q=2$, the particle spectral index $\sigma \simeq 1$, 
and for $q=5/3$, $\sigma \simeq 0.6$.

Also here the composition of the downstream wave field affects the
resulting spectrum: in the case of downstream turbulence 
calculated using the wave transmission model  
the particle population immediately behind the shock front
extends to slightly higher energies than in the equipartition case,
but for the latter the stochastic acceleration is clearly quicker. The
former effect is due to the scattering center compression ratio being larger
for lower-$M$ shocks, while the latter is due to larger velocity
differences in the equipartition wave field.


\subsection{''Re-acceleration of Re-accelerated Particles''}

\begin{figure*}
  \begin{center}
    \includegraphics[width=0.37\linewidth]{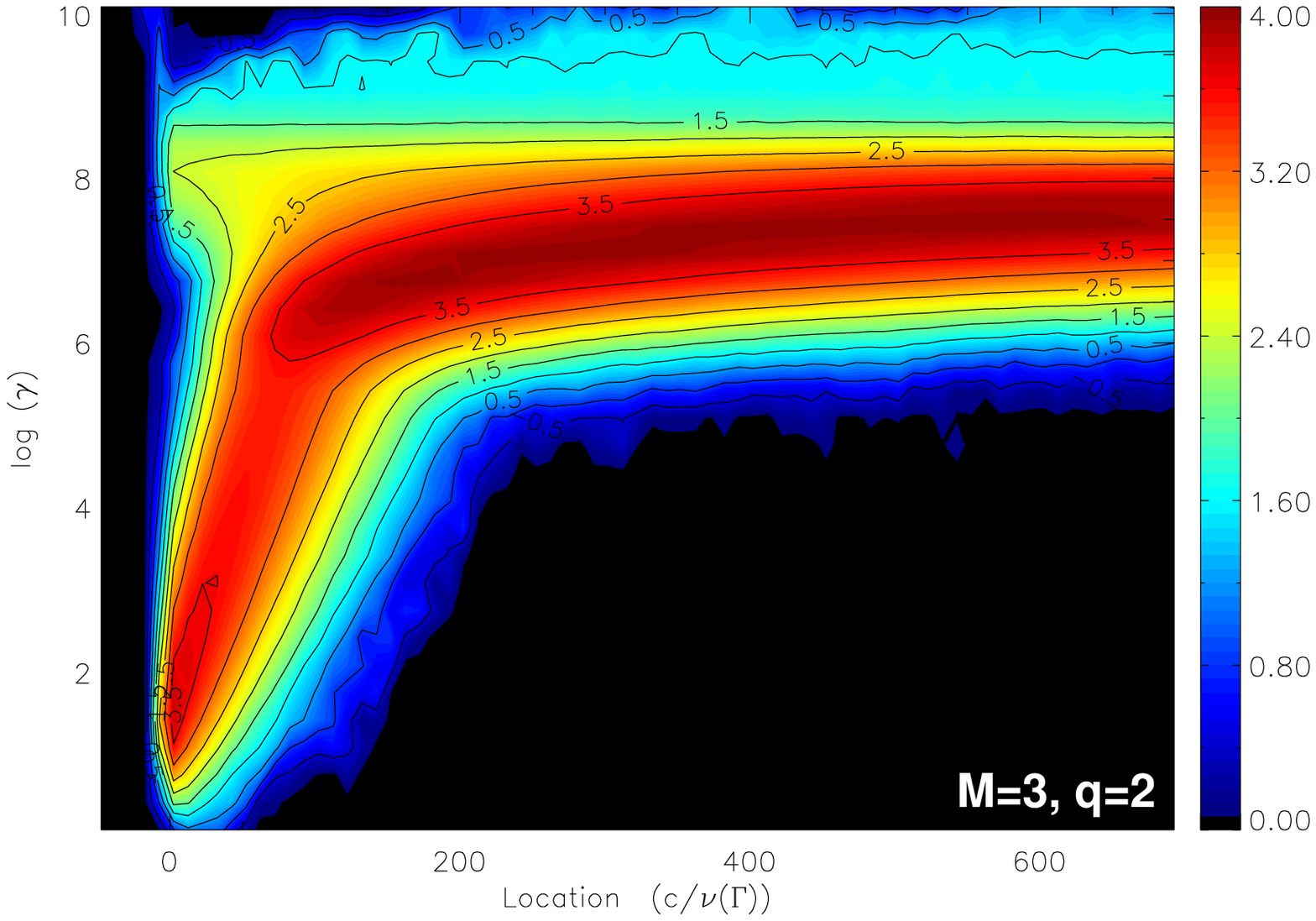}\nolinebreak 
    \includegraphics[width=0.42\linewidth]{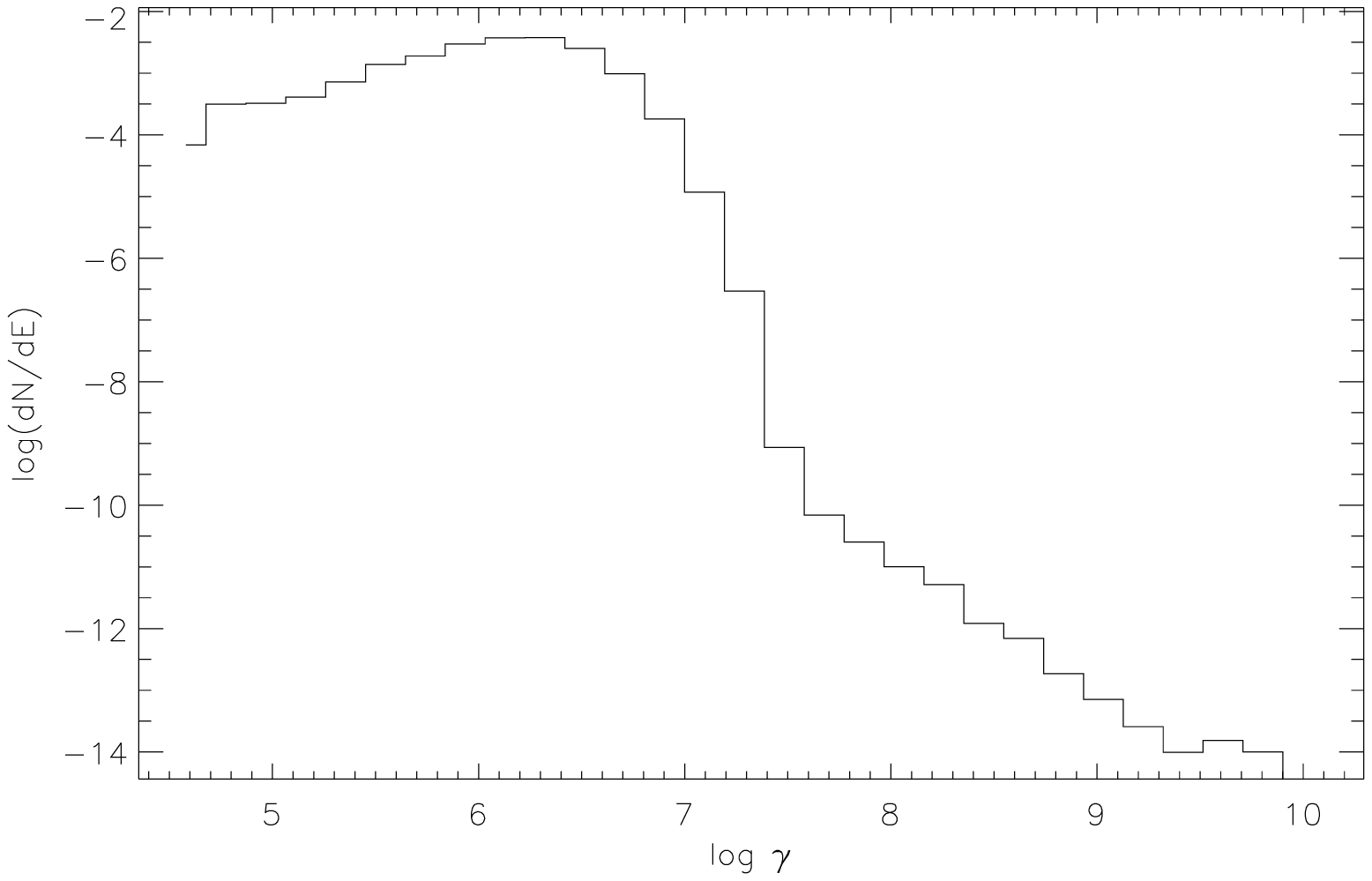}
  \end{center} 
  \caption{Stochastically accelerated particles 
    return to the shock and get re-accelerated by the 
    first-order process. On the right the particles are gathered
    at downstream free-escape boundary. \cite{VV2005}}
  \label{fig:re_acceleration}
\end{figure*}

The prolonging of the mean free path of particles with $\gamma
> \gamma_0$ (or $k_{\rm res} < k_0$) has, in addition to the drop of
the stochastic acceleration efficiency, another interesting feature:
particles that are already energized first in the shock by the
first-order mechanism and then in the downstream by the
second-order acceleration, become able to return back to the shock and get
''re-injected'' into the first-order acceleration process. Injection
energies for this second injection are very high, or course, so 
particles get accelerated to ultra-high energies; this is seen 
as a distortion of the high-energy part of the left-hand contour 
panels in Figs.~\ref{fig:contours_1} and 
\ref{fig:contours_2}, and especially clear in Fig.\ref{fig:re_acceleration},
where the low Mach number $M=3$ and turbulence corresponding to 
$q=2$, both favorable for fast stochastic acceleration, allow 
particles to be accelerated to energies of the order of $\gamma_0$ 
while still being sufficiently close to the shock front in order to 
being able to return to the first-order acceleration process. 
As a result, a clearly visible high-energy (quasi) power-law is 
seen in the upper part of the contour as well in the energy spectrum 
collected at the downstream border.


\subsection{Combination of Injection Mechanisms}
\begin{figure}
  \begin{center}
    \includegraphics[width=0.9\linewidth]{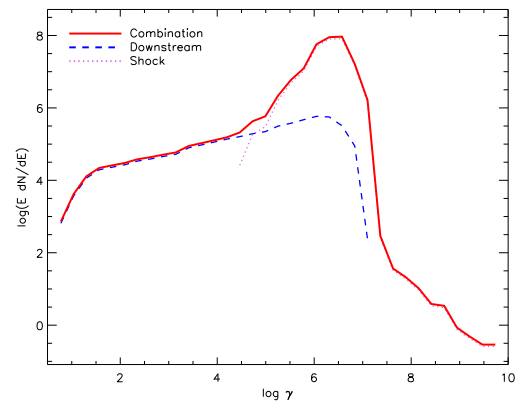}
  \end{center} 
  \caption{Example of a combination (solid line) energy spectra of
    particles injected to the acceleration process at the shock
    (dotted) and throughout the downstream region (dashed). (Model
    parameters are $q=5/3$ and $M=10$.)  The particles are collected
    at the downstream free escape boundary, $\sim 4\times10^{12}$ cm
    away from the shock, and the number of the particles injected at
    the shock is larger than the number of uniformly injected
    particles by a factor of $100$. \cite{VV2005}}
  \label{fig:combination}
\end{figure}

We also investigated what kind of particle energy spectra the two
injection schemes -- one operating at the shock, and another
operating uniformly throughout the downstream region -- are able to
create. Fig.~\ref{fig:combination} gives an example of combination 
of these. In simulations 
these two acceleration cases were mostly kept separate for the sake of
simplicity, but there should be no reason to assume the separation be
present also in nature. Also the relative amounts of shock- and
downstream-injected particles are not fixed by our model, but instead
considered a parameter.



\section{CONCLUSIONS}

We have studied stochastic particle acceleration in the downstream
region of a relativistic parallel shock. Applying the wave
transmission calculations of \cite{VVS} and assuming the cross
helicity to vanish in the upstream, we have modeled the turbulence of
the downstream region as a superposition of Alfv\'en waves propagating
parallel and anti-parallel to the plasma flow.  Using a kinetic
Monte Carlo simulation we have modeled the second-order Fermi
acceleration of electrons in the shock environment, and considered
cases of acceleration of downstream-injected particles, as well as
that of particles injected at the shock. We have shown that the
stochastic acceleration can, indeed, have remarkable effects in both
cases. This result is even more pronounced if the two downstream
Alfv\'en wave fields are assumed to be in equipartition.

The behavior of the particle energy distribution in the stochastic
process depends heavily on the strength of the background magnetic
field; in the cases of weak magnetic field and large quasi-Newtonian
Alfv\'enic Mach number the effects of stochastic acceleration are
faded out by the much stronger first order acceleration. Also the
magnetic field turbulence spectrum affects the acceleration
efficiency: for Kolmogorov turbulence with $q=5/3$ the spatial scales
are up to an order of magnitude shorter than in the case of $q=2$
turbulence.  Although the spatial scales in simulations presented here
are enormous compared to those associated with shock acceleration (the
first-order process in the immediate vicinity of the shock front), in
case of blazars and other active galactic nuclei the scales are still
orders of magnitude too small to be resolved even in the VLBI pictures
-- regardless of the turbulence and used magnetic field strength. Also
the acceleration time scales are short: the time required to
shift the whole spectrum from the initial energy range to $\gamma_{\rm
bulk} \gtrsim 10^6$ ranged from $10$ to $50$ minutes in the $M=10$
case, and for $M=3$ the times were $\lesssim 1$ minute, as measured in
the shock frame.

In the cases where the stochastic acceleration was quick enough for
particles to reach the $\gamma_0$ energy while being still
sufficiently close to the shock in order to be able to make their way
back to the upstream region due to their prolonged mean free path, the
first-order mechanism was able to re-accelerate the returning
high-energy particles to even higher energies. This led to forming of
a new (quasi-)power-law at energies $\gamma \gtrsim 10^7$ in some
cases.

One notable feature of the present model is that in the case of a
uniform injection process in the downstream region, power-law spectra
with high and low energy cut-offs are formed. Depending on the
turbulence, particle energy spectra have power-law spectral indices of
$0.5$--$1$ with lower and higher energy cut-offs at $\gamma_1 \approx
10^1 \simeq \gamma_{\rm injection}$ and $\gamma_2 \sim 10^6 =
\gamma_0$, respectively. These particles would produce synchrotron
spectra with photon spectral indices $-0.5 < \alpha < 0$ in the
GHz--THz regime for various initial parameters. These properties are
quite similar to those of flat-spectrum sources, for which typical
spectra with $\alpha \gtrsim -0.5$ in the GHz region and flare spectra
with $\alpha \approx -0.2$ in the optically thin region of the
spectrum are seen; see e.g. \cite{ValtaojaEtAl1988} and references
therein.  

To conclude, the main results of this study are: 
\begin{itemize}
\item Stochastic acceleration can be a very efficient mechanism in the
downstream region of parallel relativistic shocks, provided that the
magnetic field strength is large enough in order to make the
Alfv\'enic Mach number approach the critical Mach number ($M_{\rm c} =
\sqrt{r}$) of the shock, i.e., to increase the downstream Alfv\'en
speed enough to allow for sufficient difference in speeds of parallel
and anti-parallel Alfv\'en waves required for rapid stochastic
acceleration.
\item In the case of a continuous injection mechanism in the downstream
region particle energy distributions with hard spectral indices can 
be formed between the injection energy and $\gamma_0$. The accelerated 
particle populations could produce synchrotron spectra very similar to
those of flat spectrum sources.
\item The interplay between the first- and second-order Fermi
acceleration at relativistic shocks can produce a variety of spectral
forms not limited to single power laws.
 \end{itemize}

\noindent \textbf{Remark:} We wand to emphasize that these simulations are 
based on the test-particle
approximation, i.e., the effects of the particles on the turbulent
wave spectrum and on the shock structure are neglected. Including
these effects in a self-consistent manner may lead to notable effects
on the resulting spectrum, see e.g. \citep{Vainio2001}. 
Including these effects to our model is,
however, beyond of the scope of the present simulations.


\bigskip 

\end{document}